\documentclass[sigconf,authorversion]{acmart}

\usepackage{subcaption}
\usepackage{enumitem}
\usepackage{dirtytalk}
\usepackage{makecell}
\usepackage{multirow}

\setcopyright{rightsretained}
\copyrightyear{2026}
\acmYear{2026}
\acmConference[PEARC '26]{Practice and Experience in Advanced Research Computing}{July 26--30, 2026}{Minneapolis, MN, USA}

\begin{document}

\title{Building the Palmetto API}
\subtitle{Adding granular permissions and caching to the Slurm REST API without sacrificing compatibility}

\author{Ben Godfrey}
\affiliation{%
    \department{Research Computing and Data}
    \institution{Clemson University}
    \city{Clemson}
    \state{South Carolina}
    \country{USA}
}
\email{bfgodfr@clemson.edu}
\orcid{0009-0009-1352-2397}

\author{Doug Dawson}
\affiliation{%
    \department{Research Computing and Data}
    \institution{Clemson University}
    \city{Clemson}
    \state{South Carolina}
    \country{USA}
}
\email{dndawso@clemson.edu}
\orcid{0000-0003-3802-2531}

\begin{abstract}
The development of administrative and computational research tools requires reliable programmatic interfaces with the cluster scheduler. The Research Computing and Data (RCD) team at Clemson University has developed the Palmetto API, a proxy for the native Slurm RESTful interface, \texttt{slurmrestd}, while providing advanced authentication, authorization, and caching. This paper details the design and implementation of this proxy, evaluates the performance benefits from caching, and verifies compatibility to existing \texttt{slurmrestd} clients. The result is a light-weight and secure method of exposing our cluster scheduler to tools and automations.
\end{abstract}

\ccsdesc[500]{Information systems~Web services}
\ccsdesc[100]{Information systems~Open source software}
\ccsdesc[500]{Software and its engineering~Middleware}
\ccsdesc[100]{Security and privacy~Software and application security}

\keywords{Slurm, slurmrestd, REST, API, HTTP, web service, high-performance computing, HPC, research computing}

\received{09 February 2026}
\received[accepted]{16 March 2026}
\received[revised]{25 March 2026}

\maketitle

\section{Introduction}

Over the last several years, Clemson University's Research Computing and Data (RCD) team has worked on building and modernizing the Palmetto 2 cluster \cite{modernizing_p2}. As part of this effort, the team has built a variety of automations and integrations, including Slurm account synchronization with our ColdFront instance \cite{clemson_coldfront}, Palmetto Account Check Tool (PACT) diagnostics \cite{pact}, online hardware table, and many other internal tools. These tools need a reliable source for information about the state of the cluster, which often means getting data from the Slurm Workload Manager \cite{slurm}, our job scheduler on Palmetto 2. 

Out of the box, many Slurm installations come with \texttt{slurmrestd} \cite{slurmrestd}, which can provide a REST API for interacting with the cluster. This enables building applications and integrations that can work across many different clusters thanks to the common interface. However, it is important to note that there are security concerns when exposing \texttt{slurmrestd} to an untrusted network, as noted by the official documentation \cite{slurmrestd}, which recommends placing \texttt{slurmrestd} behind a proxy. As such, we set out to build the Palmetto API service on top of \texttt{slurmrestd} with the following goals:

\begin{enumerate}
    \item Maintain compatibility with the existing Slurm REST API specification.
    \item Add stronger authentication mechanisms with support for granular permissions.
    \item Add a caching layer to reduce load on \texttt{slurmctld}/\texttt{slurmdbd}.
    \item Create an easily extensible API service that we can add our own custom endpoints to.
\end{enumerate}

This year, the most recent release of Palmetto API has implemented all of these goals and moved into production use. Our team has spent time evaluating its impact through load testing and a compatibility survey, which yielded positive results. For other centers with similar goals, Clemson plans to release the source code of Palmetto API under the GNU Affero General Public License at the PEARC'26 conference \textbf{(link to source code will be added here)}. While some parts of the Palmetto API are inherently designed around the specific setup Clemson's HPC systems, we have made an effort to make these components configurable or at least clearly labeled. We encourage other centers to build on our implementation and make something useful for their system.

\begin{figure*}
    \includegraphics[width=\linewidth]{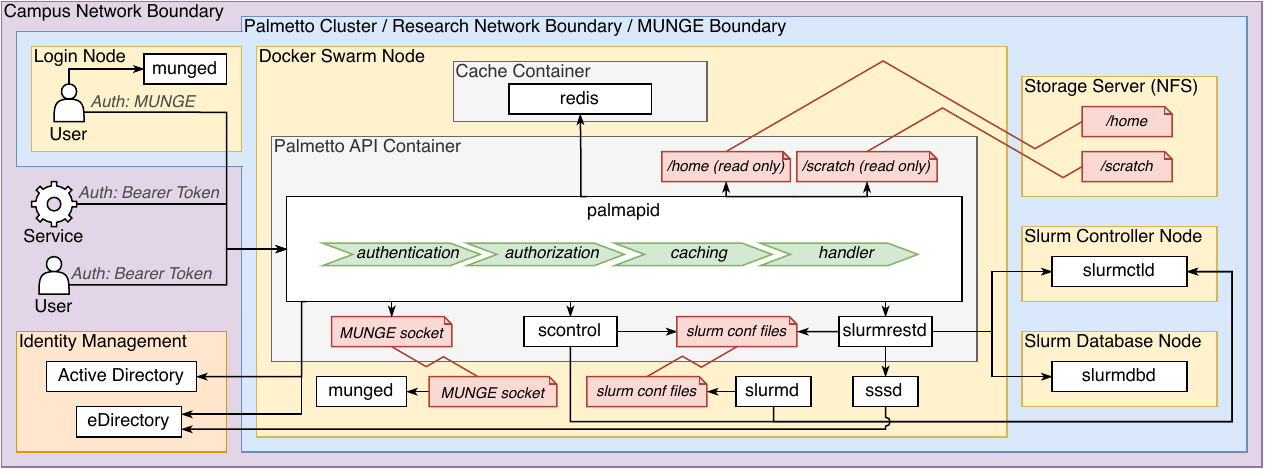}
    \caption{Palmetto API architecture diagram}
    \label{fig:arch_overview}
    \Description{
        Architecture diagram showing the many services that Palmetto API connects to across the cluster and broader campus network.
        The Palmetto API container communicates with a Redis container for caching and directory services provided by our campus identity management.
        From the host, the MUNGE socket and Slurm configuration files are mounted into the Palmetto API container.
        Home and scratch file systems are also mounted into the Palmetto API container via NFS.
        Within the container, the Palmetto API can run the scontrol and slurmrestd binaries, which communicate with slurmctld and slurmdbd over the network.
        Users (services/people) send requests to the Palmetto API service, which processes the request through its authentication, authorization, caching, and handler layers.
    }
\end{figure*}

\section{Background}

Structured, programmatic access to HPC systems provides a multitude of benefits to both HPC users and system administrators.
API access allows researchers to use complex workflow managers and build science gateways and allows administrators to connect powerful monitoring and administrative tools.

These benefits have driven many other institutions to develop API wrappers for their own HPC systems.  
For example, NERSC build their Superfacility API \cite{nersc_api} allowing job submissions to their systems and the Gesellschaft für wissenschaftliche Datenverarbeitung mbH Göttingen (GWDG) has built HPCSerA \cite{hpc_rest_auth} enabling secure API-based job submission to its HPC infrastructure.
Other institutions, including Texas Advanced Computing Center (TACC) with Tapis \cite {tapis} and the Swiss National Super Computing Centre with FirecREST \cite{firecrest} have sought to not only build API platforms for their institutional resources, but to release their API platforms as open source so that they can be used by other institutions.  
Both Tapis and FirecREST abstract the underlying details of the systems (e.g. job scheduler), allowing API consumers to submit generic jobs.
Globus Compute \cite{globus_compute} also offers a Functions-as-a-Service (FaaS) platform.  
They provide a Globus Compute Web API that handles authentication and dispatches tasks to endpoint agents HPC institutions can run on their own infrastructure.

We see the usefulness of an abstract interface for job submission for building science gateways.
However, since we were not only targeting job submission but also aiming to build powerful administrative tools, we needed to stick as close to the raw scheduler as we could.
Building a wrapper around \texttt{slurmrestd} provides a solution for our goals.  
We can build powerful administrative tools that perform account and node management, while also providing methods for users to submit and monitor their jobs through an API.
The interface is standardized and \texttt{slurmrestd} is available on most Slurm clusters, providing significant portability to clients.

\section{Architecture}

The architecture of Palmetto API involves many systems across our campus and cluster, as shown in Figure \ref{fig:arch_overview}. At the core, you will find the Palmetto API server application (\texttt{palmapid}), which is written in Go \cite{golang}. The server runs in a containerized environment, which features a Slurm installation baked into the image at build time. Other dependencies are mounted into this container at runtime from the host, which is a virtual compute node in our Slurm cluster reserved for running internal services.

In the background, \texttt{palmapid} runs several background tasks that enable it to function. These threads are responsible for maintaining persistent connections to directory services, monitoring changes from the cache, and periodically refreshing JWT tokens from Slurm. 

In the foreground, the server component of \texttt{palmapid} does the heavy lifting of answering client requests. Clients make requests to the API service via the HTTPS protocol from anywhere on the campus network. The server first passes each request through several \textbf{middleware} layers, which are responsible for authentication, authorization, and caching. If the request passes through all of the middleware, it is handed off to the associated \textbf{request handler}, which will generate the appropriate response. 

This architecture has proven to be flexible and has grown with us as we added new endpoints and features over time. Below, we explain how the most critical pieces enabled us to satisfy our goals set out earlier.

\subsection{Authentication Middleware}

Authentication middleware is one of the first security mechanisms encountered by every request to the Palmetto API. This middleware is responsible for verifying the identity of the client and determining what credentials they possess. Considering that the API serves many audiences, including users and services both on and off the cluster, this middleware must support multiple types of authentication. We accomplish this via our authentication provider interface, which makes it easy to support new methods as needed. Currently, the Palmetto API supports two authentication providers: Bearer Token and MUNGE.

\subsubsection{MUNGE Authentication}

For clients accessing the API from the cluster, we wanted an easy way to authenticate without the hassle of distributing credentials to users. Since MUNGE \cite{munge} was already being used to authenticate cluster users with Slurm, it seemed logical to support MUNGE authentication with the API as well. Clients may supply a MUNGE credential in the Authorization header of their request. 

We are able to validate MUNGE credentials on the API server using the \texttt{libmunge} library \cite{libmunge}. While this library is written in C, we can use it natively from Go via CGO \cite{cgo}. Beyond simply ensuring \texttt{munge\_decode} does not return an error, we enforce additional requirements on the credential to prevent misuse. Firstly, we require that the source IP address of the request match the host address in the MUNGE credential. Secondly, we require that the time-to-live (TTL) on the MUNGE credential be less than five seconds. 

\subsubsection{Bearer Token Authentication}

\label{sec:bearer-token-auth}

For clients accessing the API outside of the cluster, we needed an alternative authentication mechanism and decided to use bearer tokens. To create a new bearer token, our administrators generate a long unique string (the bearer value) and record it in the \texttt{tokens.toml} file along with the associated username and permissions this token should grant. When making requests, clients supply this bearer value in the \texttt{Authorization} header. The server treats bearer values as opaque and will merely check to see if it matches any known bearer token from \texttt{tokens.toml} to make an authentication decision.

Our decision to use opaque values for bearer tokens instead of a structured format like JSON Web Tokens (JWT) \cite{jwt} was primarily due to revocation. Many clients would be managed internally by our team, but we also planned to allow users to write their own integrations. Considering these clients would be outside of our control and the risk of credential theft or leaks, we needed to ensure we had the ability to deactivate tokens quickly. If we used JWT instead, due to the reliance on digital signatures for validity, we would either have to maintain a revocation list (expensive to check as it grows), rotate the JWT signing key (invalidating all tokens), or possibly use separate access and refresh tokens similar to OAuth (adding complexity) to mitigate the associated risks \cite{jwt_revocation}.

Making this token mechanism compatible with existing software required a bit of revision. Firstly, since \texttt{slurmrestd} uses the \texttt{X-Slurm-User-Token} for authentication, existing clients will expect to send their tokens in this manner. We added support for this as well so that clients can use either header. Secondly, since the current implementation of Slurm tokens is JWT, some clients may attempt to introspect this token and verify it is valid. Using a random opaque string will cause an error with these clients because the token doesn't match the expected format. As such, for compatibility, the opaque string we use for our bearer value is a JWT signed with a random key that has the same fields Slurm uses, plus an additional field with random bytes to further increase entropy beyond the signature.

\subsection{Authorization Middleware}

After the authentication phase, non-public endpoints are wrapped with authorization middlewares, called \textbf{requirements}, that verify the user is allowed to access the endpoint. Four different types of requirements are supported in Palmetto API: scope requirements, account restriction requirements, group requirements, and IP address requirements. The latter two are primarily used for administrative endpoints, with most endpoints relying on the former two, discussed below.

\subsubsection{Authorization Scopes}

Most authorization decisions in Palmetto API are based on scopes, which limit what actions a credential is allowed to perform. Each scope grants access to a specific set of related endpoints. For example, the \texttt{slurm:jobs:manage} scope allows submitting, updating, and canceling jobs. A full list of scopes and their associated privileges is listed in Appendix \ref{appendix:scopes}. These scopes are a big part of what makes Palmetto API different from a basic authenticating proxy to \texttt{slurmrestd}, as they enable fine-grained security controls while keeping things simple enough for practical use.

\subsubsection{Account Restriction Policies}

Account restriction policies limit what Slurm accounts the credential is allowed to modify. Credentials may include a list of regular expressions that match accounts it should have access to. An empty list indicates the credential has no account restrictions. Applying an account restriction policy is useful, for example, in cases where you may want to limit job submission to a certain part of your cluster or prevent a development instance from modifying production accounts. 

Enforcing account restriction policies is not a simple task, since it requires examining the request carefully to ensure only authorized accounts are involved. There were three key points that made this more feasible to maintain. First, we decided to implement account restriction policies only on mutating endpoints (methods \texttt{POST} or \texttt{DELETE}). Reading data from other accounts was not a concern for us, but we wanted to limit what actions could be taken using an account. Secondly, we separated the concern of checking which accounts were involved with the request from verifying those accounts were allowed by the credential. This allowed us to have common code for the latter, while creating separate functions that can parse each endpoint's data format for the former. Thirdly, we took advantage of the OpenAPI specification provided by Slurm \cite{slurm_oapi}, combined with the popular OpenAPI Generator \cite{oapi_gen}, to automatically generate Go structures for every endpoint and version we support. This makes it easy for us to parse request bodies and extract the account fields we need. We have already went through one Slurm upgrade with this code that included a new Slurm API version, and simply regenerating the OpenAPI client code combined with Go's strict type checking made it a breeze with minimal changes necessary.

\subsection{Caching Middleware}

To reduce load on \texttt{slurmctld} and \texttt{slurmdbd}, some endpoints are wrapped in a caching middleware. Palmetto API uses a Redis \cite{redis} instance to back its cache, configured with a least frequently used (LFU) eviction policy. Before calling the request handler, the cache middleware consults the metadata stored in Redis to see if there is a sufficiently fresh cached copy available. Only in the event of a cache miss is the request handler called to generate a new response. For each cache entry, we store the following in a Redis hash: a timestamp of when the response was generated, a timestamp of when the response is considered stale, values of HTTP headers, the HTTP response code, and a string of bytes representing the HTTP response body contents. Below, we discuss some of the challenges we encountered designing this cache and features enabled by its design.

\subsubsection{Cache Expiration Timings}

Deciding how long to cache responses for was not trivial, especially since we intended to serve cached responses by default. Some information stored in Slurm does not change very often, such as node configurations. However, some information changes frequently, such as the status and metadata of individual jobs. As a result, Palmetto API sets cache timings on a per-endpoint bases, which are set via cache policies. There are three cache policies: short (1-10 seconds), normal (10-30 seconds), and long (30-60) seconds. We have set the cache policies based on what makes sense for our cluster and intended usages, but if they do not work for you, they are easy to change. We also wanted to make the cache timings respond with cluster load. As such, we define a response's freshness lifetime based on the time it took to generate the response, adding buffer of 1-5 seconds (depending on the policy), then considering the minimum/maximum times set in the cache policy. 

\subsubsection{Cache Fallback Mechanism}

In the event that \texttt{slurmctld} or \texttt{slurmdbd} is unreachable, clients receive an error instead of the data they sought. In some cases, clients may prefer to receive the last stale response stored in the cache rather than an error. For example, there are times where our Slurm instance may be unavailable during scheduled maintenance windows. During this time, the hardware listing on our website would display an error to visitors. The cache fallback feature allows us to display the last known hardware table until Slurm returns to service.

The cache fallback mechanism is off by default, since it may cause unexpected behavior in clients that do not support it. Clients may indicate to the API service that they are okay with stale data in the event of an error via the standard \texttt{Cache-Control} header with a \texttt{stale-if-error} directive. 

To implement this feature, we had to make adjustments to how we used the built-in \texttt{EXPIRE}/time-to-live (TTL) feature of Redis, which causes keys to be deleted after a defined period. Instead of setting the TTL as the length of time the response would be fresh, we instead set the TTL to a much longer fallback expiration of 3 days. So, that means that keys are only evicted from the cache when either a) the cache is full and the LFU policy kicks in, or b) the fallback timeout of 3 days expires. Recall that we store a timestamp in each cache entry's Redis hash that indicates when it is stale, so we are still able to quickly make a freshness decision by retrieving just the timestamps as necessary.

\subsubsection{Avoiding Cache Stampeding}

With over a thousand compute nodes on Palmetto 2, we had concerns that our API may suffer from cache stampeding (also known as the thundering herd problem) in the event an automation (like our health check script) runs on many nodes at once while the cache is stale. Avoiding hundreds of simultaneous calls to Slurm for the same information when a stampede event occurs would be ideal. To accomplish this, we used the Go \texttt{singleflight} package \cite{singleflight} to group simultaneous requests for the same resource key and share a single response generation thread amongst all of them. In some of our larger experiments, this makes a big difference in the number of remote procedure calls (RPCs) sent to Slurm.

\subsubsection{Permissions and Security of Shared Responses}

The cache and response sharing system of Palmetto API was designed based on the permissions configuration of Palmetto 2. Administrators can configure the \texttt{PrivateData} setting in \texttt{slurm.conf} \cite{slurm_conf} and \texttt{slurmdbd.conf} \cite{slurmdb_conf} to control what information users can see. By default, this setting is empty and users are allowed to see everything \cite{slurm_conf, slurmdb_conf}, which is the case on Palmetto 2. In this case, there is no difference in the API response for read queries as different users, so there is no security risk with sharing responses between users. For clusters with more stringent privacy settings, it would be necessary to remove the service account from shared response generation and modify the cache key to be user specific.

\begin{table*}
    \centering

    \begin{tabular}{l|rr|rr}
        \toprule
        \multirow{2}{*}{Endpoint}                         & \multicolumn{2}{c|}{Without Cache} & \multicolumn{2}{c}{With Cache} \\
                                                          & Median          & Average          & Median        & Average        \\ \midrule
        \texttt{GET /slurmdb/v0.0.43/jobs} (last 6 hours) & 20,000 ms       & 19,336 ms        & 15,000 ms     & 15,575 ms      \\
        \texttt{GET /slurmdb/v0.0.43/users}               & 130 ms          & 134 ms           & 14 ms         & 52 ms          \\
        \texttt{GET /slurm/v0.0.43/shares}                & 3,000 ms        & 3,153 ms         & 39 ms         & 40 ms          \\
        \texttt{GET /slurm/v0.0.43/jobs}                  & 5,900 ms        & 8,033 ms         & 4,000 ms      & 2,949 ms       \\
        \texttt{GET /slurmdb/v0.0.43/associations}        & 1,300 ms        & 1,332 ms         & 76 ms         & 490 ms         \\
        \texttt{GET /slurm/v0.0.43/nodes}                 & 1,000 ms        & 1,648 ms         & 40 ms         & 141 ms         \\
        \texttt{GET /pact/self}                           & 1,800 ms        & 8,121 ms         & 250 ms        & 696 ms         \\
        \texttt{GET /hardware/stats}                      & 1,100 ms        & 1,355 ms         & 4 ms          & 63 ms          \\
        \texttt{GET /hardware/table}                      & 1,100 ms        & 1,429 ms         & 4 ms          & 37 ms          \\
        \texttt{GET /hardware/avail}                      & 960 ms          & 786 ms           & 4 ms          & 26 ms          \\ \bottomrule
    \end{tabular}

    \caption{Load test results for select endpoints, demonstrating the impact of caching.}
    \label{table:speedup}
\end{table*}

\subsection{Slurm Request Handler}

For requests for \texttt{/slurm/} and \texttt{/slurmdb/} endpoints, we have a dedicated request handler that communicates with \texttt{slurmrestd}. There are two modes of operation for \texttt{slurmrestd}: \textbf{\texttt{inetd} mode}, where a request is taken on \texttt{stdin} and the response is written to \texttt{stdout}, and \textbf{listening mode}, where \texttt{slurmrestd} acts as a server and listens on a port \cite{slurmrestd}. We decided to use \texttt{slurmrestd} in \texttt{inetd} mode so that the Slurm configuration would be reloaded with each request, which avoids having to restart the API service for changes to take effect. 

\begin{figure}
    \includegraphics[width=\linewidth]{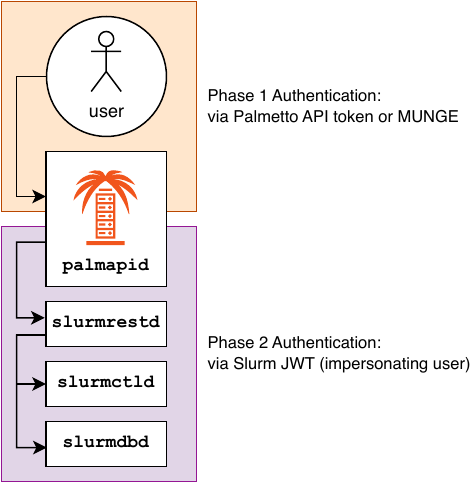}
    \caption{Diagram showing authentication methods used at each part of the connection.}
    \label{fig:auth_phases}
    \Description{
        Diagram showing the two phases of authentication that occur for a Palmetto API request.
        Phase 1 occurs between the user and Palmetto API, authenticating via Palmetto API token or MUNGE.
        Phase 2 occurs between Palmetto API, slurmrestd, and slurmctld/slurmdbd, authenticating via Slurm JWT and impersonating the user.
    }
\end{figure}

\subsubsection{Containerizing Slurm Tools}

For Palmetto API to work, it needs to execute the \texttt{scontrol} and \texttt{slurmrestd} commands, so we needed these available in the container. Our first approach was to bind-mount the binaries and their library dependencies directly into the container. The idea was to avoid building these into the container image so that we wouldn't need to update the container image if we updated Slurm. This worked well for some time, but did not work out as expected when we upgraded Slurm during a maintenance window. Since we were bind-mounting specific library files, such as \texttt{libslurm.so.42}, the API service was broken until we updated the mount to use \texttt{libslurm.so.43} for the new version of Slurm. Our second approach was to instead compile Slurm during the container build process and bake the necessary binaries and libraries directly into the container. This meant that we would need to manually update the API service when upgrading Slurm versions on the cluster. However, this ended up being the ideal approach, since Slurm maintains compatibility with prior versions \cite{slurm_upgrades}, so our API service could continue using the older clients built into the container for a brief time until we update them.

Beyond merely having the binaries and libraries available, Slurm tools also require access to read the Slurm configuration and communicate over the network with \texttt{slurmctld} and \texttt{slurmdbd}, which we would need to make work in the container. We use \say{configless} Slurm \cite{configless_slurm} on Palmetto 2, which means that the \texttt{slurmd} running on each cluster node will automatically retrieve the configuration from \texttt{slurmctld} and make it available under the \texttt{/run/slurm/conf} directory. Since the host system running Docker is a Slurm node running \texttt{slurmd} already, we can bind mount the \texttt{/run/slurm/conf} directory to make the configuration available and already know it is possible to communicate with Slurm over the network.

\subsubsection{Authenticating with Slurm}

In order to make requests on behalf of authorized users, \texttt{palmapid} needs a way to impersonate users when communicating with \texttt{slurmrestd}. In earlier revisions, prior to containerization, we relied on MUNGE authentication between \texttt{palmapid} and \texttt{slurmrestd}. This was accomplished by running \texttt{palmapid} as \texttt{root}, which allowed it to use the \texttt{setuid} and \texttt{setgid} system calls to run the \texttt{slurmrestd} process as another user. However, as we moved to expose the service to the campus network, we wanted to avoid running the service as \texttt{root} for security reasons. The most recent revisions now use JWT Authentication between \texttt{palmapid} and \texttt{slurmrestd}, which allows us to run \texttt{slurmrestd} as a service account instead. The \texttt{palmapid} service account has no privileges, meaning it does not even have an account with Slurm and cannot authenticate with Slurm on its own. In order to impersonate users with JWT authentication, we may set the \texttt{X-Slurm-User-Token} header to a JWT token for \texttt{SlurmUser} and set the \texttt{X-Slurm-User-Name} header to the username of the user we wish to impersonate \cite{slurmrestd}. A diagram explaining how this works is shown in Figure \ref{fig:auth_phases}.

Next, we needed a way to manage the \texttt{SlurmUser} JWT token. One option was to manually create a long-lived JWT and make it available as a secret in the container. This approach was simple, but meant we would need to either manually rotate this token periodically or be willing to mint a token with infinite lifetime, which we did not want to do. Instead, we decided to allow \texttt{palmapid} to generate its own \texttt{SlurmUser} JWT tokens and manage rotation automatically. This works by granting user \texttt{palmapid} permission to run \texttt{scontrol token} as the \texttt{SlurmUser} via \texttt{sudo}. Beyond this command, user \texttt{palmapid} has no other rights granted via \texttt{sudo}. A background thread will mint short-lived \texttt{SlurmUser} tokens periodically, attempting to refresh them before they expire. 

\subsubsection{Making Requests to \texttt{slurmrestd}}

Putting together all of the puzzle pieces above, we can now make requests to \texttt{slurmrestd} from within the container. The Slurm request handler of the API works in tandem with our \texttt{slurmrestd} HTTP transport function to generate and parse a response. The following procedure is used:

\begin{enumerate}
    \item Since \texttt{slurmrestd} does not support HTTP/2, rewrite requests as HTTP/1.1 if necessary.
    \item Strip all request headers, except \texttt{Accept}, \texttt{Content-Length}, and \texttt{Content-Type}.
    \item Fetch the latest \texttt{SlurmUser} JWT token from the token manager and set the \texttt{X-Slurm-User-Token} header.
    \item Read the request context to determine the calling user and set the \texttt{X-Slurm-User-Name} header to their username.
    \item Serialize the request to its wire representation.
    \item Execute \texttt{slurmrestd}, piping the wire representation to \texttt{stdin}.
    \item After \texttt{slurmrestd} exits, parse the data received on \texttt{stdout} as an HTTP response.
\end{enumerate}

At the end of this process, we return this HTTP response to the client.

\section{Evaluation}

Once we had the new containerized version of the Palmetto API up and running, we first worked on integrating it with our in-house tools, then moved on to evaluating how it may function at scale and with existing software. We performed load testing and a compatibility survey to measure this and evaluate whether we met our goals.

\subsection{Load Testing}

To evaluate the performance benefit of our cache, we performed load testing against the API using Locust \cite{locust}. We ran the load test against the API twice, first with the cache disabled and then with the cache enabled. Each test used the same procedure and ran with 24 worker nodes on our cluster making requests for 10 minutes. Results for selected endpoints are shown in Table \ref{table:speedup}. Across the board, it is easy to see the drastic impact that caching has on response times. Beyond the response timings, there is also a noticeable decrease in the error rate. Without the cache and response sharing layer, the Slurm controller and database struggle to keep up with the load and we observed timeout errors on several requests. Thanks to the cache reducing how often we must make requests and response sharing ensuring only one simultaneous request is made for the same data, we are able to reduce the number of RPCs that the Slurm controller must answer, which results in a higher success rate for the necessary RPCs.

\subsection{Compatibility Survey}

We also wanted to evaluate how successful Palmetto API is at maintaining compatibility with tools aimed at \texttt{slurmrestd}.
To do this, we found several open source projects that advertised support for \texttt{slurmrestd}, and we tested pointing them at Palmetto API.

The results are summarized in Table \ref{table:compat}.  
The example projects all eventually worked, however some required changes to the code.
The simplest example, using the \texttt{openapi-generator-cli} to generate a Python client \cite{openapi_gen_python}, worked with no changes.
The Slurm HPC Dashboard \cite{slurmdash} required a minor change to the upstream project and we introduced a pull request that allows custom ports (rather than hard-coded 6820).
We also had to make a change to how we minted bearer tokens for the Palmetto API.
We realized that some \texttt{slurmrestd} clients, including the Slurm HPC Dashboard project, introspect the provided authenticating JWT to look for expiration rather than treat it as an opaque value.
Instead of generating random opaque bearer tokens, we began to mint JWTs with the same claims that Slurm uses (see Section \ref{sec:bearer-token-auth}).

The last two projects (a Slurm Prometheus exporter \cite{slurm_exporter} and the Slurm Monitor website \cite{slurm_monitor}) both required API version changes to be compatible with our current cluster.
The Slurm exporter could be updated easily to target a newer API version as the structure of the responses didn't change in relevant ways between single version bump needed to be compatible.
Updating the Slurm Monitor proved to be more time consuming.  
It required updating the version from \texttt{v0.0.38} to \texttt{v0.0.41} and there were breaking changes to fields used in the response.
The effort needed to update was also increased due to the fact that the project did not use an OpenAPI generator and did not use type hinting, so there was no automatic build time method of checking whether changing the API version was a breaking change or not.

Based on our experiences with the selected projects, we believe the Palmetto API should be a drop in replacement for most clients that expect to interact with \texttt{slurmrestd}.  
The most likely cause for incompatibility could be if the project uses a deprecated API version.
Unfortunately, we expect this to be pretty common since Slurm only supports a few API versions in each release meaning clients have to constantly update \cite{slurm_oapi}.
A client a few years older than the Slurm cluster will likely be incompatible.
Furthermore, we make the following recommendations for client authors:

\begin{itemize}
    \item Allow flexibility in URL and Token.
    \item Use OpenAPI generators to generate the interface code.
    \item Use strongly typed languages or type hinting to detect when API version changes break the code base.
\end{itemize}

\begin{table}
    \centering
    \begin{tabular}{
        p{0.30\columnwidth}
        p{0.62\columnwidth}
    }
        \toprule
        \thead{Program}  & \thead{Notes} \\
        \midrule
        OpenAPI Generated Python \cite{openapi_gen_python} &
                No changes needed. 
                \\
        \midrule
        Slurm HPC Dashboard \cite{slurmdash} &
                We submitted a patch to upstream for custom custom port number support.
                \newline
                Tokens must be real JWTs.
                \\
        \midrule
        slurm-exporter \cite{slurm_exporter} &
                Project expects to be able to fetch Slurm token with \texttt{scontrol}. 
                \newline
                API version needed to be updated from v0.0.40 to v0.0.41.
                \\
        \midrule
        slurm-monitor \cite{slurm_monitor} &
                API needed to be updated version from v0.0.38 to v0.0.41, including data structure changes.
                \newline
                Hard-coded URL needed to be updated.
                \\
        \bottomrule
    \end{tabular}
    \caption{Lists programs using slurmrestd tested for compatibility with Palmetto API and the result.}
    \label{table:compat}
\end{table}

\section{Future Work}

At this stage, we are pleased with the stability of the API, but there is always room for improvements. Firstly, we would like to add support for Globus Auth \cite{globus_auth} as a third authentication provider for the API. As we evaluate ways for science gateways from our institution's researchers to connect with the cluster, possibly via the Palmetto API, we want to make it easy for others to register their own clients, which Globus makes it easy for developers to do. The growing list of institutions supported by Globus' identity federation also might make this auth provider useful for other centers. Secondly, we would like to make this project easier to adapt for use with other clusters. The initial release has some configuration, but there are many areas where this could be expanded. We also recognize that adding feature flags to disable Clemson-specific features would also be helpful.

\section{Conclusion}

After over a year of development and now several months in production, we have had some time to reflect over the Palmetto API design and our goals. In particular, the compatibility promise has been both a challenge and a blessing. We knew we wanted compatibility from the outset, but had to reconcile with how our additions on top of the existing interface would affect existing clients. Fortunately, we did not find very many compatibility issues in the end with our testing and look forward to automatically reaping the benefits of future improvements to the \texttt{slurmrestd} underneath our project. Secondly, we have enjoyed the granular permissions when testing third-party apps. Being confident that we know what actions other software can take on our cluster via the API has enabled us to test more community applications and begin to allow our researchers to build their own integrations. Thirdly, the caching layer has been valuable for reducing load against our Slurm controller and database. The cache is largely transparent to clients, as we intended, but having the metadata and control headers available has ensured pickier clients can still make the right choices. On a cluster with over a thousand nodes, not having to worry that a simultaneous run of the health check script will overfill the RPC queue is liberating for our administrators. Lastly, we have enjoyed having a solid foundation to extend upon and add new functionality. We have already rewritten our hardware table integration as a custom API endpoint, adding to the existing library of custom endpoints, like the PACT diagnostics tool.

We are excited to release the Palmetto API as open source software to the community at PEARC'26 under the GNU Affero General Public License \textbf{(link to source code will be added here)}. This initial release creates a solid base to work from and our team's documentation of our experience building it can serve as a guide for others. In future releases, we hope to make this project easier to adapt for use on other clusters so it can be used by more centers. 


\bibliographystyle{ACM-Reference-Format}
\bibliography{mendeley}

\appendix

\section{Authorization Scopes}\label{appendix:scopes}

\begin{description} 
    \item[\texttt{slurm:read}] read data from the Slurm controller.
    \item[\texttt{slurm:jobs:manage}] submit, update, or cancel jobs.
    \item[\texttt{slurm:nodes:manage}] add, update, or remove nodes.
    \item[\texttt{slurm:reservations:manage}] create, update, or delete resource reservations.
    \item[\texttt{slurm:reconfigure}] ask the controller to reconfigure.
    \item[\texttt{slurmdb:read}] read data from the Slurm database.
    \item[\texttt{slurmdb:accounts:manage}] create, update, or delete accounts.
    \item[\texttt{slurmdb:accounts\_association:manage}] create, update, or \\ delete accounts associations.
    \item[\texttt{slurmdb:associations:manage}] create, update, or delete associations.
    \item[\texttt{slurmdb:clusters:manage}] create, update, or delete clusters.
    \item[\texttt{slurmdb:config:read}] read the database configuration.
    \item[\texttt{slurmdb:config:manage}] update the database configuration.
    \item[\texttt{slurmdb:qos:manage}] create, update, or delete quality of service instances.
    \item[\texttt{slurmdb:users:manage}] create, update, or delete users.
    \item[\texttt{slurmdb:users\_association:manage}] create, update, or delete users associations.
    \item[\texttt{slurmdb:tres:manage}] create, update, or delete trackable resources.
    \item[\texttt{slurmdb:wckeys:manage}] create, update, or delete work characterization keys.
    \item[\texttt{pact:self}] run PACT \cite{pact} assessments against your own account.
    \item[\texttt{pact:admin}] run PACT \cite{pact} assessments against any user account.
    \item[\texttt{health:check}] check the health of Palmetto API components.
    \item[\texttt{pprof:read}] read data from the Go performance profiling system.
\end{description}

\end{document}